\documentclass{article}
\pdfoutput=1
\usepackage{jcappub}

\newcommand{\be}{\begin{equation}}
\newcommand{\ee}{\end{equation}}
\newcommand{\bea}{\begin{eqnarray}}
\newcommand{\eea}{\end{eqnarray}}
\newcommand{\eq}{\begin{equation}}
\newcommand{\eqx}{\end{equation}}
\newcommand{\eqn}{\begin{eqnarray}}
\newcommand{\bi}{\begin{itemize}}
\newcommand{\eqnx}{\end{eqnarray}}
\newcommand{\ei}{\end{itemize}}
\newcounter{hran}

\newcommand{\ba}{\begin{array}}
\newcommand{\ea}{\end{array}}
\newcommand{\balg}{\begin{align}}
\newcommand{\ealg}{\end{align}}

\renewcommand{\thefigure}{\Roman{figure}}

\newcommand{\lsim}
{\raise0.3ex\hbox{$\;<$\kern-0.75em\raise-1.1ex\hbox{$\sim\;$}}}
\newcommand{\gsim}
{\raise0.3ex\hbox{$\;>$\kern-0.75em\raise-1.1ex\hbox{$\sim\;$}}}

\title{Phenomenology of the minimal inflation scenario: inflationary trajectories and particle production}
\author[a]{Luis \'Alvarez-Gaum\'e,}
\author[b]{C\'esar G\'omez,} 
\author[c]{and Raul Jimenez}

\affiliation[a]{Theory Group, Physics Department, CERN, CH-1211, Geneva 23, Switzerland}
\affiliation[b]{Physics Department and Instituto de Fisica Teorica UAM/CSIC, 28049 Cantoblanco, Madrid, Spain}
\affiliation[c]{ICREA \& ICC, Universitat de Barcelona (IEEC-UB), Marti i Franques 1, Barcelona 08028, Spain}

\emailAdd{luis.alvarez-gaume@cern.ch}
\emailAdd{cesar.gomez@uam.es}
\emailAdd{raul.jimenez@icc.ub.edu}

\abstract{We discuss the phenomenology of the minimal inflation scenario. We concentrate on 
two aspects: inflationary trajectories and particle 
production. Our findings can be summarized in two main results: first, that inflationary 
trayectories that are very flat and provide enough number 
of e-foldings are natural in the scenario without fine tunning. 
We present a general formalism  to
identify attractors in multi-field inflation  regardless of trajectories fulfilling the slow-roll conditions. 
We then explore particle production in
the model and show how the inflaton naturally transmutes into a dark matter particle. One interesting feature of our model is 
that it provides a novel mechanism to generate particles and entropy in the universe: the filling of the Fermi 
sphere up to a given momentum $p_F$ due to the sea of goldstinos that are an important part of the
matter generated after inflation. With this mechanism in hand we predict that the 
gravitino should have a mass of $> 100-1000$ TeV.  Another interesting feature of our model is that the predicted level of gravity waves is $r=0.1 - 0.001$, which is in the range of detectability from Planck and upcoming CMB polarization experiments.}

\begin{document} 

\maketitle

\section{Introduction}

In recent realizations of the theory of inflation\cite{inflation}, it is most natural to use multi-field inflation, 
that is the inflaton field is a composite of
many fields, with some recent examples where the number of fields can be very large \cite{wands}. It is 
difficult to think of examples of the 
inflaton based on fundamental physics models where one can be limited to only a single field. While it can 
be the case that many of the fields 
are ``frozen" and that only one of them drives the inflationary period, it is most often the case that several fields participate in the inflationary
period. In order to determine the existence of attractors in multi-field inflation it is common practice, in order to determine if a given 
trajectory produces enough number of e-foldings to explain the size of the current universe, to impose the so-called slow-roll condition {\it 
simultaneously} in all fields that drive inflation \cite{sasakistewart}. This is, of course, a sufficient but not a necessary condition. It 
remains an open problem in multi-field inflation to determine the attractor in phase-space. Further, in multifield inflation the inflaton trajectories can be dominated in part by non-linear friction terms. Motivated by an interest in the existence of 
the attractor trajectories in multi-field inflation and to study the phenomenology of the minimal inflation scenario \cite{minfI,minfII}  we 
have decided to revisit the definition of  "slow-roll" conditions in multi-field models of inflation. 

We have also investigated particle production in the minimal inflation scenario in order to explore the fate of the inflaton. The reason we have 
done this, and not used standard (pre)(re)-heating calculations for a scalar field,  is that in some minimal inflation scenarios the inflaton is not a 
fundamental field but a bilinear fermion, or a combination of a fermion bilinear and a scalar field.
Therefore, under certain conditions the inflaton field can ``transmute" in part into a sea of fermions and end up 
as part of the constituents of the Universe. 

In the minimal-inflation\cite{minfI,minfII} scenario the field $X$ that drives the exponential expansion of the Universe can often be represented at low
 energies by a Goldstino composite $GG$. Our main motivation to propose to identify the inflaton field with the order parameter of supersymmetry 
 breaking is guided by the fact that, independently of the particular microscopic mechanism driving supersymmetry breaking (in what follows we will
 restrict ourselves to $F$-breaking ) we can define a superfield $X$ whose $\theta$ 
 component at large distances becomes the "Goldstino" (see \cite{volkovakulovrocek,seiberg2}).  In the UV the scalar component $x$ of $X$ is well
 defined as a fundamental field while in the IR, once supersymmetry is spontaneously broken, this scalar field may be expressed as a
two Goldstino state.  The explicit realisation of $x$ as a fermion bilinear depends on the low-energy details of the model.
In models of low-energy supersymmetry the realization of $x$ as $GG$ can be 
implemented by imposing a non linear constraint in the IR for the $X$ field of the type 
$X^2=0$. In our previous approach to inflation we used one real component of the UV $x$ field as the
inflaton. We assumed the existence of a F-breaking effective superpotential for the $X$-superfield 
and we induced a potential for $x$ from gravitational corrections to the K\"ahler potential.  In the cosmological
context we have to be careful to verify if the conditions for the decoupling of the scalar field $x$, or rather, its
transmutation into a  goldstino bilinear are realised after the system exits inflation and begins the pre-heating 
period.  In the simplest models analysed below this condition is not completely satisfied, hence, after inflation,
we obtain a universe filled with a fermi gas (liquid) generated by the goldstinos together with a massive scalar
component. 

It is this property of this model of inflaton that allows us to use the inflaton as 
the generator of dark matter. When the inflaton abandons the slow-roll conditions ($\eta > 1$; where $\eta$ is the second slow-roll parameter), we
can work out the pre-heating and reheating of the universe. We have thus a possible candidate for
dark matter.  Since our model only depends on the scale of supersymmetry breaking, we get tight 
constraints for the expected mass of the dark matter particles ($\sim 100$ TeV). 

\section{Setup}

In multi--field inflation the question of finding the attractor for inflationary 
trajectories is more complicated than with a single field.  In the case of 
caothic inflation  the attractor can be found analytically 
\cite{Mukhanov:2005sc}. Here we describe a general algorithm to obtain attractors in multi-field inflation.

With a flat FLRW ansatz for the space-time metric, we consider the equations of motion for a generic
model.  The scalar fields will take values on some target space manifold with metric $g_{ij}(\Phi)$.
The first FLRW equation relates the Hubble parameter to the energy density:
\begin{equation}
H^2 = \frac{8 \pi G}{3} \left ( \frac{1}{2} g_{ij} \dot \Phi^i \dot \Phi^j + V(\Phi) \right ),
\end{equation}
where the $\Phi^i$  are the scalar fields of the theory, with the inflaton among them, and 
we have the usual definitions for pressure ($p$)  and energy density ($\rho$)
\begin{equation}
p =  \frac{1}{2} g_{ij} \dot \Phi^i \dot \Phi^j - V(\Phi); \,\,\, \rho = \frac{1}{2} g_{ij} \dot \Phi^i \dot \Phi^j + V(\Phi).
\end{equation}

We also have the second cosmological equation:
\begin{equation}
\frac{\ddot a}{a} = - \frac{4 \pi G}{3} \left ( \rho + 3p \right )
\end{equation}

To have inflation, we require that $p = - \rho$, as this will give an exponential 
growth of the scale factor $a = \exp(\int H(t) dt)$. Thus we need to find trajectories with 
$p \sim - \rho$ for enough time: more than 55 e-folds to explain the current universe.
The equation of motion for the scalar field reads:
\begin{equation}
\frac{D \dot \Phi^i}{dt} + 3 H \dot \Phi^i + g^{ij} V_j (\Phi) = 0
\end{equation}
where $D$ is the covariant derivative with respect to the target space metric $g_{ij}$.

In our case, the scalar fields form a complex scalar field, the partner of the goldstino field.  Although we
will make explicit the arguments below in this case, the arguments easily generalise to a more complex situation.
Our complex field can be written as $z = M (\alpha + i \beta)/\sqrt{2}$ (hereafter M is the Planck mass scale). 
The potential is $V(z, \bar z) = f^2 V(\alpha, \beta)$, since we only include for simplicity two scales, 
the Planck scale M and the  supersymmetry breaking scale $f^{1/2}$.  In supergravity models, the gravitino
mass is up to simple numerical factors given by $m_{3/2}\sim f/M$.
It is convenient to write down dimensionless equations of motion such that time 
is counted in units of $f^{-1/2}$.  If $L^3$ is 
the space-time volume, then the Kahler metric reads
\begin{equation}
ds^2 = 2 g_{z \bar z} ds d \bar z = \partial_z \partial_{\bar z} K(\alpha, \beta) M^2 (d\alpha^2 + d\beta^2).
\end{equation}
The action is then:
\begin{equation}
S = L^3 \int dt a^3 \left (\frac{1}{2} g(\alpha, \beta) M^2 (\dot \alpha^2 + \dot\beta^2) - f^2 V(\alpha, \beta) \right),
\end{equation}
where $t = \tau \,M/f $, thus
\begin{equation}
S = L^3 f^2 m_{3/2}^{-1} \int d\tau a^3 \left (\frac{1}{2} g(\alpha, \beta)  (\alpha'^2 + \beta'^2) - V(\alpha, \beta) \right).
\end{equation}

Hence the system of differential equations for the trayectory becomes:
\begin{eqnarray}
\alpha'' + 3 \frac{a'}{a} \alpha' + \frac{1}{2} \partial_{\alpha} \log g (\alpha'^2 - \beta'^2) + \partial_{\beta} \log g \alpha' \beta' + g^{-1} V'_{\alpha}  & = & 0 \nonumber \\
\beta'' + 3 \frac{a'}{a} \beta' + \frac{1}{2} \partial_{\beta} \log g (\beta'^2 - \alpha'^2) + \partial_{\alpha} \log g \alpha' \beta' + g^{-1} V'_{\beta} & = & 0 \nonumber \\
\frac{a'}{a} = \frac{H}{m_{3/2}} = \frac{1}{\sqrt{3}} \left ( \frac{1}{2} g  (\alpha'^2 + \beta'^2) + V(\alpha, \beta) \right ) &  &
\end{eqnarray}
Next we need to understand the conditions leading to slow roll, and also the corresponding basin of attraction
in the space of initial conditions leading to them. In general there is no need to start with conditions where
the scalar field is at rest.  They will evolve in time to get closer and closer to the slow roll attractor.
To determine the attractor we notice that the slow roll conditions imply that in (2.4) we can ignore the
acceleration term $D \dot \Phi^i / dt \sim 0$ which represents geodesics in the target space.  This implies
that the trajectory is dominated by the lower derivatives terms 
$\alpha' = -g^{-1} V_{\alpha}/(3 H), \beta' = -g^{-1} V_{\beta}/(3 H)$.  The initial conditions for the trajectories
are then provided by these equations at some fiducial initial time which yields $\alpha'(0),\beta'(0)$, but 
with the Hubble parameter $H$ determined by (2.1).  Replacing the derivatives determined by the first order
equations in (2.1) yields a fourth order equation for $H$.  The only acceptable solution is:
\begin{equation}
H = \sqrt \frac{1}{18}  \left ( 3 V +  \sqrt{6 V' + 9 V^2}  \right )
\end{equation}
We now have the relevant initial conditions for the geodesic equation $D \dot \Phi^i / dt \sim 0$.  We
start with the same initial conditions for the geodesic equation and the first order equations representing
slow roll, and analyse how long the trajectories stay together.  This is shown in the right panel of Fig.~\ref{fig:1}. So long as the geodesics and the exact solution to the equations of motion remain close enough, the 
slow roll conditions hold.  This argument should hold for the general case.  Rather than looking
for general analytic criteria, we now look for a numerical sampling of the initial conditions to
find the attractor trajectories.  To illustrate the existence of inflationary 
trajectories and attractors in phase space for multi-field inflation we focus on two examples 
in our minimal inflation scenario: the massles case and the "half-tube" case to be explained below.

\section{Some examples of minimal inflation potentials with very flat directions}

In our minimal inflationary scenario \cite{minfI,minfII} we use only the Ferrara-Zumino (FZ)
multiplet to drive inflation.  The scalar potential in the Eisntein frame is given by:
\begin{equation}
V_{E} = e^{\frac{K}{M^2}}( -\frac{3}{M^2}W \bar W +G^{X\bar X} D_XW \, D_{\bar X}\bar W),
\end{equation}
where the K\"ahler metric and the K\"ahler covariant derivatives are given by:
\begin{equation}
G_{X\bar X}\,=\,\partial_X \bar \partial_X\, K(X,\bar X)\qquad
D\,W(X)\,=\,\partial_X\, W(X)\,+\,{\frac{1} {M^2}}\,\partial_X\,K\, W(X)
\end{equation}
In our approach we make an explicit, but resaonably generic choice for $K$ and $W$.

For us the inflaton superfield is the FZ-chiral superfield $X = z +\sqrt{2}\,\theta\psi + \theta^2 F$,
the order parameter of supersymmetry breaking. We will consider the simplest superpotential implementing
F-breaking of supersymmetry.  More elaborate superpotentials often reduce to this one once heavy fields
are integrated out.
\begin{equation}
W = f X + f_0\, M 
\end{equation}
with $f_0$ some constant to be fixed later by imposing the existence of a global minimum with 
vanishing cosmological constant and with $f$ the supersymmetry breaking scale $f= \mu_{susy}^2$.

We are interested in subplanckian inflation, and not on the ultraviolet complete theory
that should underlie the scenario.  Hence we simply parametrize the subplanckian theory
in terms of the previous superpotential, and a general K\"ahler potential whose coefficient
will be taken of order one.  We try to use our ignorance of the ultraviolet theory to
our advantage. The K\"ahler potential $K$ we consider is:
\begin{equation}
K= X \bar X + \frac{a}{2 M} (X^2\,\bar X + c.c.) - \frac{b}{6 M^2}(X \bar X)^2
-\frac{c}{9 M^2} (X^3 \bar X + c.c.)+\ldots\,-2\,M^2\,
\log(1 + \frac{X +\bar X}{M})
\end{equation}

The most straightforward interpretation of why we write this K\"ahler potential is that we 
consider all terms up-to $1/M^2$ power and a log term explicitly breaking all possible R-symmetries. 
The coefficients $f_0, a, b, c$ will be chosen appropriately so that we obtain flat directions. 
For our purposes it is convenient to chose $a=0$, as this guarantees the existence of a global 
minimum. $f_0$ will be adjusted such that  the global minimum is at a vanishing value
of the potential. So the model only has $b,c$ 
as free parameters, which we try to keep of order one to avoid fine tunning in the potential. 

From the collection of potentials considered, not all will show flat directions 
where it is possible to inflate during enough of e-foldings ($> 55$) for any choice of the 
microscopic parameters $f_0, a, b, c$. We will restrict to cases where the potential has a global 
minimum with vanishing cosmological constant and thus we fix 
the value of the minimum at $0$ tuning the value of $f_0$.

\begin{figure}
\begin{center}
\includegraphics[width=.6\columnwidth]{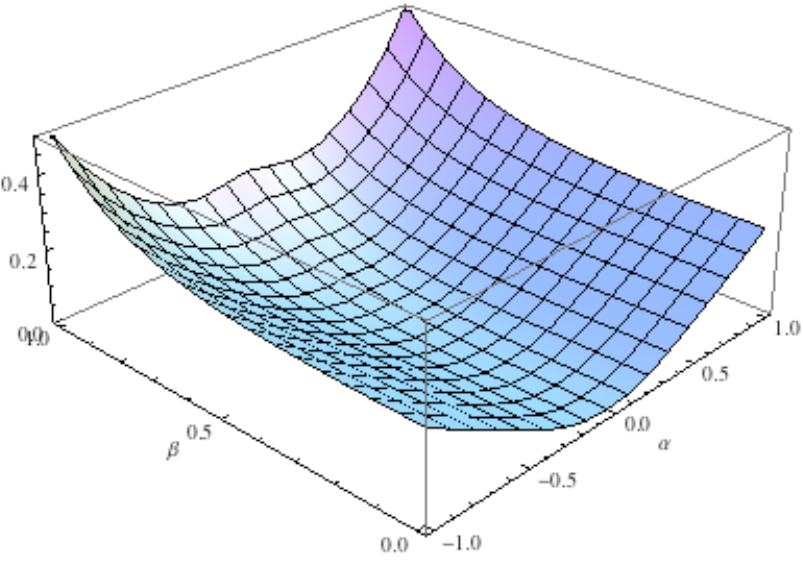}
\includegraphics[width=.35\columnwidth,height=12cm]{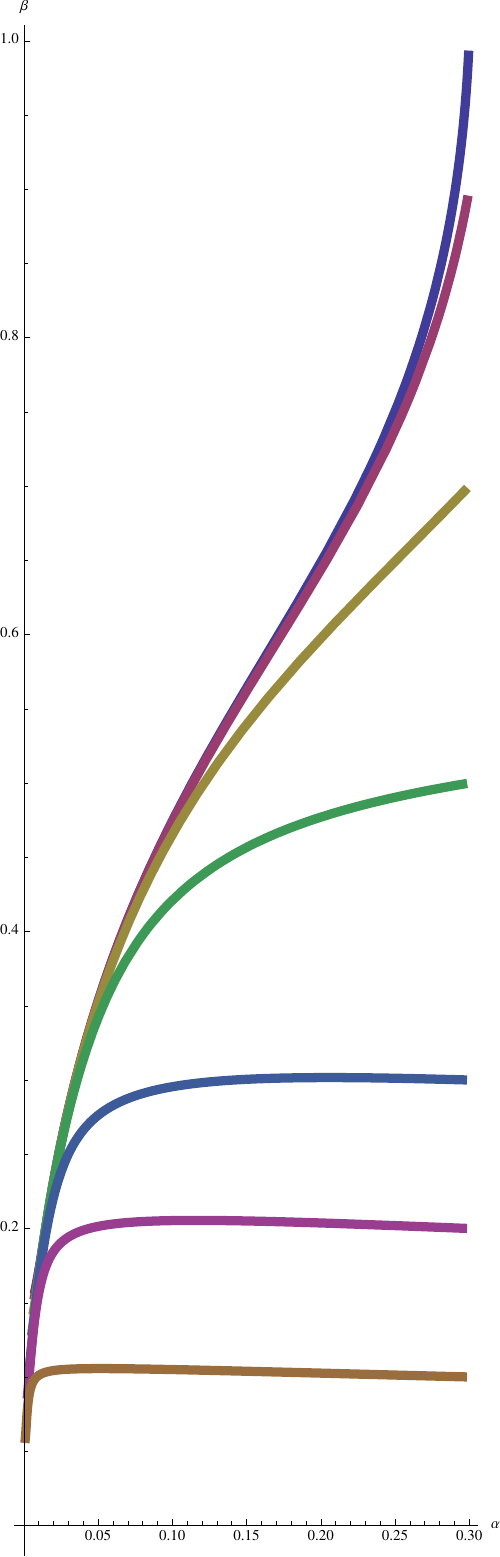}
\end{center}
\caption{Left-panel: the potential (shown in units of $V/f^2$) for the massless case as a function of the real fields $\alpha,\beta$. Right-panel: inflationary trajectories computed with the formalism described in the text to choose initial conditions. Note the existence of an attractor that brings all the trajectories to the flat region of the potential.}
\label{fig:1}
\end{figure}

For simplicity consider first the massless case, where $f_0 = -f$, $a=0, b=1, c=0$. This is the
simplest minimal model as it has only one parameter $b$, which we set to $1$ for concreteness. 
The potential has the shape shown in the left 
panel of Fig.~\ref{fig:1}. The obvious flat direction runs parallel to the $\beta$ direction. If we solve the equations 
of motion, the slow-roll trajectories look like those shown in the right panel of Fig.~\ref{fig:1}, where we have chosen arbitrary 
points to start the field. Note that all trajectories flow toward an attractor parallel to the $\beta$ direction, 
i.e. the one that is quite flat and therefore the one providing enough e-foldings.

\begin{figure}
\begin{center}
\includegraphics[width=.45\columnwidth]{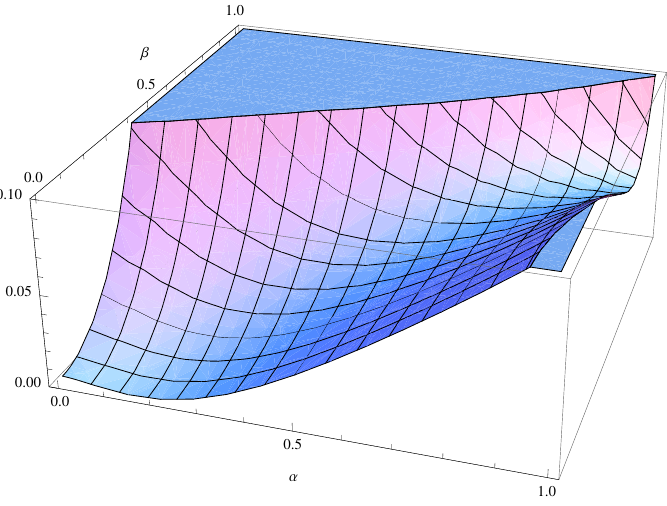}
\includegraphics[width=.45\columnwidth]{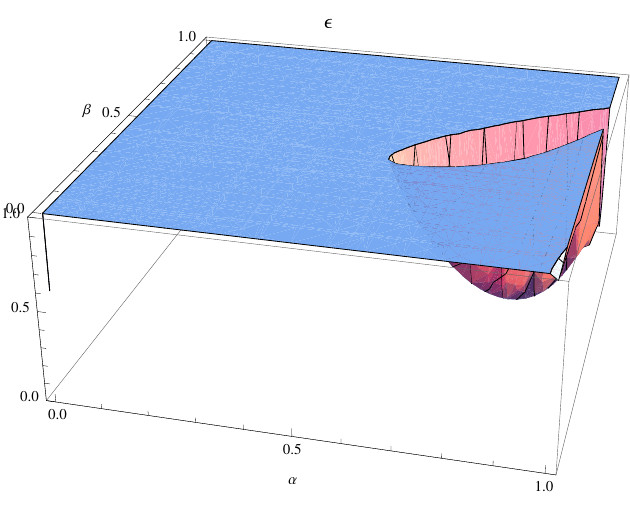}
\includegraphics[width=.45\columnwidth]{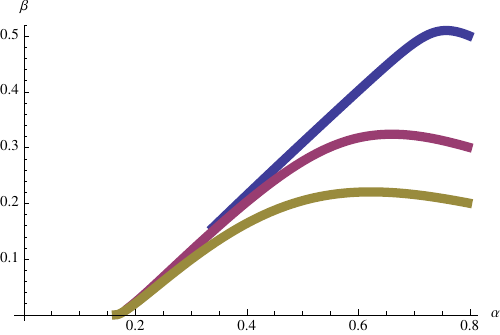}
\includegraphics[width=.35\columnwidth]{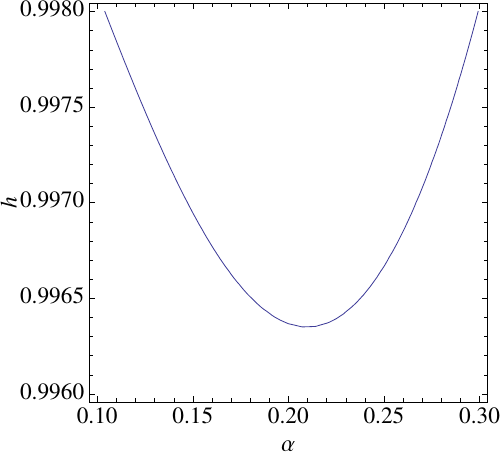}
\hspace*{-3.5cm}
\includegraphics[width=.45\columnwidth,height=4cm]{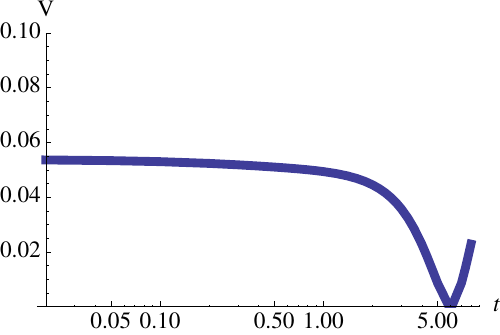}
\end{center}
\caption{Upper-left panel: The potential (shown in units of $V/f^2$) as a function of the real fields ($\alpha,\beta$) for the case with parameters ($a = 0, b = 1, c = -1.7, f_0 \sim -f$, note that this creates a very flat "bottom-valley" region, where slow-roll conditions are fulfilled. The fact that this region is very flat is shown in the upper-right panel, 
where we show the value of the first slow-roll parameter added in the two real directions. Trajectories for arbitrary 
initial conditions, as chosen by the algorithm described in the text, are shown in the middle-left panel.  Note that all 
converge to an attractor around the flat region. Middle-right panel: the minimisation of $h$ as a function of $\alpha$ 
in order to determine the minimum of the potential with zero value (or to match it to the present value of the 
cosmological constant). Bottom-left panel: the value of the potential as a function of time $t$ (as defined in text) 
for the blue trajectory; the flat region is clearly visible as well as the exit from slow-roll, which starts particle production, yielding a canonical inflationary behavior}
\label{fig:2}
\end{figure}

\begin{figure}
\begin{center}
\includegraphics[width=\columnwidth]{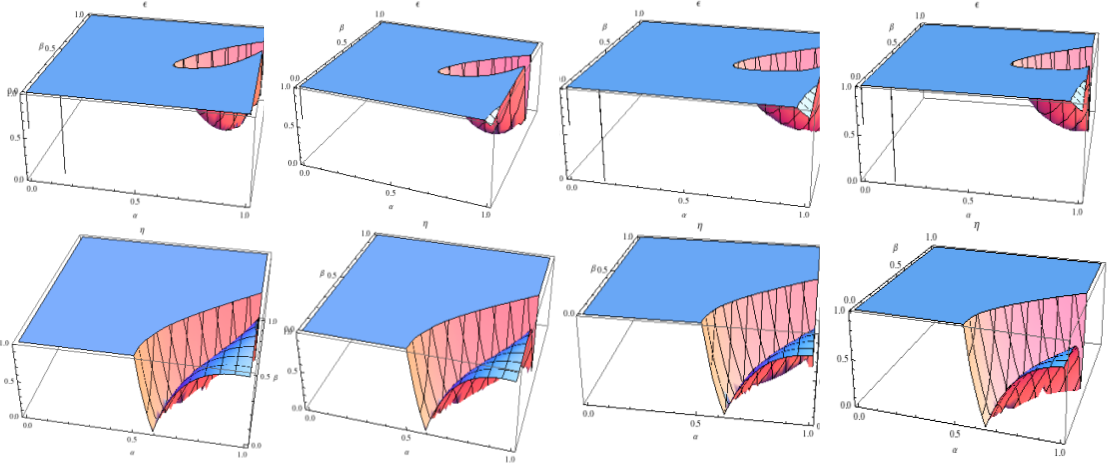}
\end{center}
\caption{The traditional slow-roll parameters ($\epsilon, \eta$) are computed for different values of the potential in the ``half-pipe'' potential show that in all cases both $\epsilon, \eta < 1$. The values of the potential are chosen randomly from a monte-carlo realization of $1000$ trials that keeps those values that fullfil the slow-roll conditions. The values, from left-to-right, are: $(a=0, b=1, c=-1.7)$,  $(a=0, b=1, c=-1.65)$, $(a=0, b=1.1, c=-1.65)$, $(a=0, b=1.15, c=-1.7)$.}
\label{fig:3}
\end{figure}

A very interesting case of potentials with very flat directions are those where $a=0$ and $c = - g \times b$, where 
g is a coefficient order one. This provides a very flat direction along a "half-pipe" with the added advantage 
that the steep directions provide a large set of initial conditions leading to inflation. 
In these models a global minimum 
always exits. The value of the potential at this minimum is not necessarily zero by simply 
choosing $f_0 = -f$, so $f_0$ needs to be chosen differently. To do this 
let us define $f_0 = -h f$. Since the potential is symmetric in $\beta$ and the global minimum, for 
these models, is always at $\beta =0$, we only need to minimise the potential for values of $h, g, \alpha$. 
Because $g$ is order one in all cases, the potential will depend only on $h,\alpha$, or rather the dependence
on $g$ is very weak. Therefore, the equation to 
find $h,\alpha$ will be $V(h,\alpha) =0$ (see left-bottom panel of Fig.~\ref{fig:2}). This equation has many solutions, 
but only the one for which $\frac{\partial h}{\partial \alpha} = 0$ will provide a solution with a global 
minimum equal to zero. This can be done by solving the following implicit equation (where the dependence in 
$g$ has been dropped because it is very weak)

\begin{equation}
-\frac{3}{4} \left(-2 h+\sqrt{2} \alpha \right)^2+\frac{\left(1-\frac{\left(2 h-\sqrt{2} \alpha \right) \left(-72+\alpha  \left(\sqrt{2}+2 \alpha \right) \left(18+5 \alpha ^2\right)\right)}{72 \left(1+\sqrt{2} \alpha \right)}\right)^2}{1+\frac{\alpha ^2}{3}+\frac{2}{\left(1+\sqrt{2} \alpha \right)^2}} = 0
 \end{equation}
 
 A fitting formula providing an accurate solution is $f_0 = -1 + 1.2 \times 10^{-3} g$. This provides 
 the full solution to find values for $f_0, b, c$ (recall that we always keep $a=0$) where the global minimum 
 is at $0$ and a very flat direction exist. Let us study some examples.

 If we choose $a=0, b=1, c=-1.7, f_0 = -.99819156 f$ then we obtain the potential shown in the upper-left 
 panel of  Fig.~\ref{fig:2}. This potential has a very flat direction for sub-planckian values 
 in both directions $\alpha$ and $\beta$. This can be easily checked by computing the first 
 slow-roll parameter in both directions and adding them $\epsilon = g^{ij} V,i V,j/V^2$ (upper-right panel of Fig.~\ref{fig:2}). 
 This is of course a sufficient {\em but not necessary} condition for inflation. Wherever this number is smaller 
 than one there is inflation (see also Fig.~\ref{fig:3}). 

We can now compute trajectories and attractor in more detail using the equations from the previous 
section. This is shown in the lower-right panel of Fig.~\ref{fig:2}. As expected all trajectories pass through 
the flat region, therefore any inflationary trajectory provides sufficient number of e-foldings. 
The bottom left panel of Fig.~\ref{fig:2} shows the value of the potential along the inflationary trajectory
as a function of time.  

The level of gravitational waves is easily predicted in our model. Recall that $r \approx (V/3.3. \times 10^{16} {\rm GeV})^4$. In our case, for inflationary trajectories that reproduce the amplitude of the fluctuations of the CMB, $V \sim f^2$, where $\sqrt f$ is in the range $10^{13-15}$ GeV.  Therefore $r \sim 0.1-0.001$, which is in the range of detectability from Planck and upcoming CMB polarization experiments.

\section{Particle production at the end of inflation.}

After inflation ends, energy and entropy in the form of radiation need to be produced in order 
to describe the observed universe. One interesting feature of our choice of the field $X$ as the 
inflaton, is the fact that because it is a bilinear fermion, at low energies it will manifest 
itself as a fermion rather than a scalar. There is then the possibility that gravitino production 
can be achieved directly without the need of parametric reheating. This is the scenario we explore in this section.

When supersymmetry is spontaneously broken, we can follow the flow of $X$ to the 
infrared (IR). In the deep IR this field satisfies a nonlinear constraint and becomes the ÒgoldstinoÓ superfield.

\begin{equation}
X^2_{\rm NL} = 0, 
\end{equation}

\begin{equation}
X_{\rm NL} = \frac{G^2}{2F} + \sqrt 2 \theta G + \theta^2 F 
\end{equation}

The scalar component $x$ of $X$ becomes a goldstino bilinear. Its
fermionic component is the goldstino fermion $G$, and $F$ is the
auxiliary field that gets the vacuum expectation value. This nonlinear constraint
is a consequence of integrating out the $x$-field.  To do this,
the effective inflaton mass needs to be larger than the mass of the gravitino.
To be more precise, $m_{\rm INF} > 2 m_{3/2}$ would be 
the required condition for the light fields to be integrated out. When we talk about integrating 
out we always refer to the above non-linear constraint being realised (see e.g. \cite{seiberg2}). 
Ref.~\cite{seiberg2} have shown that this non-linear constraint occurs always in Minkowski space. 
Although, after the inflationary period one is in curved space (FLRW), it is not difficult to see 
that corrections to the Minkowski result ($X^2 = 0$) will come as powers of $1/M^2$, and thus 
corrections will be suppressed for sub-plankian fields, as is the case of our model.
The masses of the inflaton and the gravitino depend on the parameters of the model.  In
minimal models like the ones considered in this paper there is no room to have complete
decoupling of the scalar field, which is only realised partially, hence the final state
after inflation will contain a mixture of a Fermi liquid state in fermions together with
a gas of inflatons.

Let us compute the masses of the inflaton in the real vacuum for the "half-pipe" model. 
The masses in the two real directions $\alpha, \beta$ can be well approximated by the following formula 

\begin{equation}
m_{\rm INF} = 0.5  \frac{f}{M}
\end{equation}

This is very similar to the gravitino mass $m_{3/2} = f/M$, 
which is the mass obtained for 
the gravitino in most models of supersymmetry breaking. We can be 
more precise and compute the gravitino mass 
as $m_{3/2} = |W| e ^\frac{K}{2 M^2}/\sqrt{3}$. For several values of the coefficients $a, b, c$ 
this gives values of $m_{3/2} = (0.2-0.5) \frac{f}{M}$. So this reinforces the fact that both masses 
are very similar and that the non-linear constraint is 
 not fully realised. Conversion of $X$ into $GG$ will take
 place as soon as the inflaton field abandons the slow-roll condition, 
 where its effective mass is zero, and adquires an effective mass. If we 
 assume that the most favorable scenario takes place we can 
 immediately see that the Fermi sphere will be filled up to 
 momentum $p_F = \sqrt{\frac{f}{M} m_{\rm INF}}$. We can now 
 estimate a lower bound for the gravitino mass to be able to produce enough of them at a rate that overcomes the expansion of the universe-recall that we are now in a FLRW phase. The observed number of particles in the universe is $n_{\chi} \sim 10^{70-90}$. Using the computed value for the Fermi momentum, this implies that  $\sqrt{f} > 10^{12} GeV$ which gives a lower bound on the gravitino mass: $m_{\chi} > 100-1000$ TeV.

However, as we have seen above, because the similarities 
of the mass of both the inflaton and goldstino fields, 
it does not seem to be a very efficient route of 
gravitino production, even though some conversion will 
take place. We have then to explore in more conventional terms 
particle production. This is done following standard arguments,
and we will not dwell on them here.  We will come back to a more
realistic treatment (including creation or ordinary matter, not
just gravitinos and inflatons in a future publication).
The basic mechanism to produce particles after inflation
is to use broad parametric resonances via oscillations of the fields 
around its minimum (see for instance Ref.~\cite{Mukhanov:2005sc} and references therein).
For the fermions, we can follow the arguments in Ref.~\cite{Giudice:1999yt,Kallosh:1999jj}. 

The simplest way to proceed is to first choose the unitary gauge in the 
full supergravity lagrangian, and given that we are working below the Planck
scale, we can safely ignore the quartic gravitino coupling.  The only
quartic terms that one should keep are those associated to the sigma
model associated to the rigid supersymmetric sigma-model representing
the naive $M\rightarrow \infty$ limit.  The relevant coupling of the
$X$-field to the gravitino is through the term $W e ^\frac{K}{2 M^2}
\bar{\psi_a}\sigma^{ab} \psi_b/\sqrt{3}$.  Although the problem seems
complicated, it can be shown \cite{Giudice:1999yt} that we can decompose the problem in
two parts:  the production of the spin-$3/2$ components of the gravitino
that is highly suppressed, and the production of the spin-$1/2$ component
(effectively the absorbed goldstino given the gravitino a mass), which
is not suppressed by powers of $M$.  Furthermore, after some manipulations
the authors of Ref.~\cite{Giudice:1999yt} show that the problem is formally identical with the
production of ordinary spin-$1/2$ fermions coupled to the inflaton field.
The lagrangian takes the form:
\begin{equation}
{\cal L} = e ^{K/2 M^2}\,\frac{f_0+f\,X}{M^2} \bar\Psi\,\Psi,
\end{equation}
for all practical purposes the contribution of the K-prefactor
is negligible. 
We can now use results from the literature to compute the particule production rate
\cite{Kofman:1997yn,Dolgov:1989us,Greene:1998nh,Greene:2000ew}.
In brief, they quantize the field following the standard semi-classical approximation to 
quantize fermions (Bogolyubov which in turn leads to a  Hartree-Fock eq.). The important result is that the 
eigenmode solution obeys an oscillator-like equation with complex frequency,
then one proceeds to solve this equation to find the particle density in an expanding background.
In the case of a flat universe, we can separate variables, each mode is characterised by
a given momentum $k$, and to a first approximation we end up with a Fermi sphere in FLRW.
The interesting result in Ref.~\cite{Greene:2000ew} 
is that only very few (2-5) oscillations produce values of $n_k$ of order a few tenths 
(remember that because we are dealing with fermions $n_k$ has to be below $1/2$). 
Even for just two oscillations  $n_k \sim 0.1$. One can see that the Fermi sphere fills rather 
quickly, with
$p_F = \sqrt{\frac{f}{M} m_{\rm INF}}$. Which requiring 
again that the Fermi sphere fills to a level such that we produce enough particles 
($\sim 10^{70-90}$) gives a limit on the mass of the gravitino $> 1000-10000$ TeV.  Here
we are not considering the contribution to the matter content of the inflaton field.
In the type of models presented, the $X$-field does not fully decouple.  The ideal scenario
would be the one where at the end of inflation, the inflaton "integrates" itself out.  this
would require its mass to be sufficiently higher that the gravitino mass, so that it can
be safely integrate out, and then we enter the non-linear regime where soon after inflation
we have a Fermi gas of goldstinos in the FLRW universe.  This of course requires considering
more elaborate models, where one should also include the possibility of producing ordinary
matter (squarks, quarks, leptons etc) that we have so far ignored systematically.  

There are however some results from the literature that hint toward the fact that baryons can be produced in the model (see e.g. the review in Ref.~\cite{Allahverdi:2010xz}). To be more specific, there is however a couple of routes one could explore to achieve the desire production of baryons, or more precisely to excite solely the MSSM degrees of freedom: one could argue that inflation always ends in a vacuum where all the minimal super-symetric model (MSSM) gauge group is recovered; Êone advantage of doing so is to embed inflation completely within a visible sector, i.e. MSSM inflation models where inflation occurs via gauge invariant F-flat directions. Alternatively,  one would have Êto know how inflaton couples to all the hidden sectors and the visible sector. In certain examples, such as in the case of large volume compactifications, where due to hierarchy in scales and the fact that SUSY is broken by fluxes, one can actually minimize the hidden sectors to some extent. We hope
to come back to these issues in a future publication.

Finally, we want to conclude remarking the fact that one firm prediction of our model is that the  SUSY breaking scale has to be in the range $10^{13-15}$ GeV, which in turns gives  large masses for the gravitino (using that $m_{\rm grav} = f / M$).  In order to determine the fate of the gravitino as a possible component of the dark matter we will investigate in a future publication how our model interacts with the MSSM (or similar realizations) to also explore how baryons are being produced. This requires knowledge of the microscopic details.

\section*{Acknowledgements}
We thank A. Achucarro, P. Gonzalez \& M. Postma  for encouraging us to investigate 
in detail the inflationary trajectories in our model and A. Mazumdar for interesting discussions on re-heating. We warmly thank an anonymous referee for useful comments and suggestions that improved the manuscript. RJ acknowledges discussions with 
C. Germani on the roll of friction in facilitating slow-roll conditions for small-field models.

%
%
\end{document}